\documentstyle[preprint,aps]{revtex}

\begin{document}
\newcommand\ie {{\it i.e. }}
\newcommand\eg {{\it e.g. }}
\newcommand\etc{{\it etc. }}
\newcommand\cf {{\it cf.  }}
\newcommand\etal {{\it et al. }}

\newcommand{\be}{\begin{eqnarray}}
\newcommand{\ee}{\end{eqnarray}}

\newcommand{\xf}{$x_F$}
\newcommand{\jp}{$ J/ \psi $}
\newcommand{\pp}{$ \psi^{ \prime} $}
\newcommand{\ppp}{$ \psi^{ \prime \prime } $}
\newcommand{\upsln}{$ \Upsilon $}
\newcommand{\ups}{$ \Upsilon_{1S} $}
\newcommand{\upsp}{$ \Upsilon_{2S+3S} $}
\newcommand{\ccbar}{$ c \bar c$}

\newcommand{\dd}[2]{$ #1 \overline #2 $}
\newcommand\noi {\noindent}

\preprint{
	CCAST-95-02
}
\title{Phenomenology of \xf~Dependence of Quarkonium Production in
Proton-Nucleus Interactions
	\footnote{\it Partly supported by the National Science Foundation of
        China}}
\author{W. Q. Chao$^{1,2,3}$, C. S. Gao$^{1,3,4}$ and Y. B. He$^{1,4}$}
\address{ 1. China Center of Advanced Science and Technology (World
Laboratory),
       \\P.O. Box 8730, Beijing 100080,  China\\
	2. Institute of High Energy Physics, Academia Sinica, P.O. Box
	918(4),	Beijing 100039, China\\
	3. Institute of Theoretical Physics, Academia Sinica, P.O. Box
	2735, Beijing 100080, China\\
	4. Physics Department, Peking University,
         Beijing 100871, China}
\maketitle
\vfill

\begin{abstract}
  We present a phenomenological study of the $x_F$ dependence of quarkonium
production in high energy proton-nucleus collisions. The \xf~dependence of
comover contributions is introduced to account for the observed
quarkonium suppression at low $x_F$.
Combining comover contributions, nuclear shadowing effect, energy loss
mechanism
and nuclear absorption together we reproduce the overall \xf~dependence of
E772/E789 data.
\end{abstract}

\narrowtext
\newpage
\section{Introduction}

A strong suppression of charmonium ($ J/ \psi $, \pp) and bottomonium (\ups,
\upsp) production has been observed in $p-A$ collisions on heavy relative
to light nuclear targets \cite{bad83,kar90,ald91,lei92}.
Special attentions have
been paid to the significant suppressions at low (small or negative) and
high $x_F$  regions.
The suppression at low $x_F$  is usually believed
to come from nuclear absorption and
comover contributions.
The maximum contribution of nuclear absorption
to the \jp~suppression was studied in ref.\cite{gav94},
and it was found that nuclear absorption alone cannot account for the
\jp~suppression measured in either $p-A$ or $A-B$ collisions.
Since at E772 energy most of the
physical bound states are formed well outside target nucleus due to the
Lorentz dilation of the formation time,
nuclear absorption can contribute little to the observed suppression
at low \xf~region.
While comovers can continue to interact well outside
of the nuclear volume, they may in principle play a more important role in
proton-nucleus collisions,
although the number of comovers in $p-A$ reactions is not so large as in
$A-B$ collisions.
In the literature
comover contributions are generally estimated in the central
rapidity region without \xf~dependence.
To our knowledge there is no direct explanation
of the E772/E789 data at $x_F \simeq 0$ and
below \cite{ald91,lei92}. In this work we use a simple model to extrapolate
the rapidity distribution of comovers for $p-A$ reactions from $p-p$ data
\cite{tho77}, introducing the $x_F$ dependence of comover contributions,
to explain the E772/E789 data.

On the other hand, several mechanisms have been proposed to account for the
suppression at large $x_F$ . Among these, nuclear shadowing effect alone is
argued to be unable to explain the data \cite{ald91}. The intrinsic charm
model \cite{bro89,vog91} was expected to explain this $x_F$ dependence.
However, E789 data of $ J/ \psi $ production at very large $x_F$ \cite{kow94}
show
no evidence of intrinsic charm contribution. The energy loss mechanism
\cite{gav92} could account for the observed suppression at large $x_F$ .
Nevertheless, the amount of energy
loss needed to explain the E772 $ J/ \psi $ data
appears to be significantly larger than the value determined from other
considerations \cite{bro93}.
In our work, we combine energy loss mechanism together with nuclear
shadowing effect, nuclear absorption and comover contributions, and reproduce
the large \xf~data with a smaller amount of energy loss.

\section{Comover contributions}
In the case of quarkonium production, e.g., \jp~production, in proton-nucleus
collisions, comoving secondary particles can scatter with the \ccbar~pair,
and contribute to the observed \jp~suppression. These comover contributions
have been studied in the literature \cite{gav88,vog88,gav90,vog91}.
However, since
the shape of the rapidity density of the comovers
in inclusive \jp~production for p-A
collisions is unknown, comover density is usually estimated in the
central rapidity region, and taken as a constant independent of
the comover rapidity. In this work we will use
a simple model to extrapolate the rapidity distribution of comovers for
p-A reactions from p-p data \cite{tho77}. Assuming that comovers with rapidity
close to the rapidity of a \ccbar~pair can cause the breakup of the \ccbar~
pair, we translate the rapidity distribution of comovers into \xf~distribution,
and include the resulting \xf~dependence of comover contributions to explain
the E772/E789 low \xf~data.
We will follow the work of refs.\cite{vog91,gav90},
and improve the treatment of comover contributions.

Including only the comover contributions,
the $x_F$ dependent cross section of $ J/ \psi $ production in $p-A$
collisions can be expressed as
\be
{d\sigma^{pA} \over dx_F} = {d\sigma^{pp} \over dx_F}\int d^2 b
\int\limits_{-\infty}^{+\infty}dz \rho_A(b,z)
exp\left\{-\int_{\tau_0}^{\tau_f}d\tau \sigma_{co} v n(\tau,b)\right\},
\ee
where $\rho_A$ is the nuclear density profile,
$\sigma_{co}$ is the (\ccbar)-comover
absorption cross section, $\tau_0$ is the formation time of comovers,
$\tau_f$ is the effective proper time over which the comovers can interact
with $c \bar c$ pair, $v$ is the relative velocity of $c \bar c$
with the comovers, and
$n(\tau,b)$ is the density of comovers at the proper time $\tau$ and
impact parameter $b$.

Assuming Bjorken's hydrodynamics \cite{bjo83}, the comover density varies
with the proper time according to,
\be
n(\tau)={\tau_0 n(\tau_0) \over \tau}.
\ee

The comovers consist mostly of $\pi$, $\rho$ and $\omega$ mesons. We relate
$n$, the density of comovers, to
$n_{ch}$, the density of produced charged particles, by $n=f n_{ch}$.
For example, $f \simeq 1.5$ if comovers are all
$\pi$'s with equal numbers of $\pi^{+}$, $\pi^{-}$ and $\pi^{0}$.

The charged particle density at $\tau_0$ can be related to rapidity
distribution by
\be
n_{ch}(\tau_0,b)={1 \over \sigma_{in}\tau_0} {dN_{ch}^{pA}(b) \over dy},
\ee
where $\sigma_{in} \simeq 30~mb$ is the nucleon-nucleon inelastic scattering
cross section.

Now we use the so called {\it independent cluster} model \cite{cha83} to
extrapolate the charged particle rapidity distribution for $p-A$ reactions
from $p-p$ data \cite{tho77}.

We fit the data at $\sqrt{s}=45.2~ GeV$, which is close to the E772 energy,
 in the form,
\be
{dN_{ch}^{pp} \over d\eta} = Wexp\left\{-(\eta+\eta_0)^2/ \delta\right\}
+Wexp\left\{-(\eta-\eta_0)^2/ \delta\right\}
\ee
with $W$=1.75, $\eta_0$=1.5 and $\delta$=2.9.

In the framework of the independent cluster model, we have
\be
{dN_{ch}^{pA}(b) \over d\eta} = \nu (b)Wexp\left\{-(\eta+\eta_0)^2/ \delta
\right\}
+Wexp\left\{-(\eta-\eta_0)^2/ \delta\right\},
\ee
where $\nu(b)=\sigma_{in} \int dz \rho_A(b,z)$ is the number of target
participants at the impact parameter $b$.

In order to introduce the $x_F$ dependence of comover contributions, we
assume that those comovers with rapidity close to the rapidity of a
$ J/ \psi $ can cause the breakup of the $ J/ \psi $, and we relate the
rapidity of
$ J/ \psi $ in the center-of-mass system to $x_F$, which is defined as
$x_F=P_L^{*}/(P_L^{*})_{max}$, by
\be
y^{*}={1 \over 2}ln\left({E^{*}+P_L^{*} \over E^{*}-P_L^{*}}\right),
\ee
where $E^{*}$ and $P_L^{*}$ are the energy and longitudinal
momentum of $ J/ \psi $ in the c.m.s..

In our approach to the \xf~dependent comover
density two main parameters are $f$ and $\tau_0$. In Fig.1 we show
the \xf~dependence of comover densities for various sets of $f$ and
$\tau_0$. Note that the dotted curve with $f=1.5 \times 0.6$ and
$\tau_0=0.8fm$ is plotted in view of the argument of ref.\cite{gav88},
which represents a comover density too low to explain the data.
In our further calculations we will fix $f=1.5$. The comover formation
time has been estimated to be about $0.4-1.2~fm/c$, the time scale of
both quark and hadron formation in the early stages of the collision
\cite{bro88,wong}. It can be seen from Fig.1 that the comover density
is sensitive to the choice of $\tau_0$.
In the next section we will vary $\tau_0$ and $\sigma_{co}$ to attribute
the observed suppression at low \xf~ region to comover contributions.

Since comovers can interact with the produced heavy quarks $Q$ or $\bar Q$
before the quarkonium bound state can be formed, and the interaction is
independent of whether or not the heavy quark pair is produced as a color
octet or singlet, we expect that for charmonium production the (\ccbar)-comover
 interaction cross section is the same for the \jp~and \pp~bound
states. Furthermore, the comover contributions are expected to be
independent of the final hadronic size. As a result,
although we expect the occurrence of significant comover absorptions with
strong nulear dependence, the \jp~and \pp~
production would have  the same nuclear dependence under comover
absorption, which is consistent with the
earlier experimental observation \cite{ald91}.

\section{Combination of mechanisms, numerical results and discussions}
In this section we combine comover contributions, nuclear absorption, nuclear
shadowing effect and energy loss mechanism to reproduce the overall \xf~
dependence of E772/E789 data on \jp~and \upsln~production.

We will calculate \xf~dependence
of $\alpha$ for \jp~and \upsln~production in the parametrization,
\be
{d\sigma^{pA} \over dx_F} = {d\sigma^{pp} \over dx_F}A^{\alpha(x_F)}.
\ee
For \jp~production, we make use of the fact that about $60\%$ of the final
\jp 's are produced directly, about $30\%$ come from $\chi_c$ intermediate
states and the remaining $10\%$ are produced through the decay of \pp.
Analogously, about $41\%$ of the final \ups 's are produced directly,
about $35\%$ and $18\%$ come from the decay of $\chi_b$ and $\chi_b^{\prime}$,
about $5\%$ and $1\%$ are produced through the decay of $\Upsilon_{2S}$
and $\Upsilon_{3S}$ \cite{pil93}.

The cross section of quarkonium production in p-A collisions is
\be
\label{crosseq}
\nonumber
{d\sigma^{pA} \over dx_F} = {d\sigma^{pp^{*}} \over dx_F}\int d^2 b
\int\limits_{-\infty}^{+\infty}dz \rho_A(b,z)exp\left\{-\int\limits_z^{+\infty}
dz^{\prime}\rho_A(b,z^{\prime})\sigma_{abs}(z^{\prime}-z)\right\}
\\
\times
exp\left\{-\int_{\tau_0}^{\tau_f}d\tau \sigma_{co} v n(\tau,b)\right\}
\ee
where several mechanisms have been included.

The averaged nucleon-nucleon cross section of $p-A$ reactions,
$\sigma^{pp^{*}}$ in eq.~(\ref{crosseq}), is the same as the bare
nucleon-nucleon cross section except that in $\sigma^{pp^{*}}$
the distribution functions
of the gluon, quark and antiquark within the target nucleus
are nuclear
modified.
We follow the work of ref.\cite{qiu87} to incorporate the shadowing effect
into  $\sigma^{pp^{*}}$, and use
parameters determined in ref.\cite{vog91}.

The first exponential in eq.~(\ref{crosseq}) represents the contribution of
nuclear absorption. We take the values of parameters in the following way:
the formation time of charmonia and bottomonia,
$\tau_{\psi}=0.89~fm$, $\tau_{\Upsilon}=0.76~fm$, \etc,
is taken from ref.\cite{kar88},
the absorption cross section between resonances and nucleon is chosen
as
\be
\sigma_{RN}=\sigma_{\psi N}(r_R/r_{\psi})^2,
\ee
with $\sigma_{\psi N}=6~mb$, and the radii of
resonances $r_R$ taken from ref.\cite{kar88}.
In particular, we want to point out
that with the reason discussed in the first section,
the nuclear absorption can contribute very little suppression even with a
rather large absorption cross section $\sigma_{\psi N}=6~mb$,
as  indicated
by the fact that the nuclear dependence of \jp~and \pp~production is the same
within errors \cite{ald91}.

We follow the work of ref.\cite{gav92} to include the contribution
of energy loss mechanism in our calculations. In ref.\cite{gav92}
 only energy
loss mechanism is taken into account, and an upper bound of energy loss is
obtained, $dE/dZ \simeq 1.5~ GeV/fm$. Moreover, the energy loss contribution in
ref.\cite{gav92} has not been normalized, which implies an integrated
suppression. In our calculations, we complete the normalization in a similar
way
as in ref.\cite{kha93},
\be
{d \sigma ^{pA} \over dx_F}(x_F)\mid_{normal.}={1 \over \alpha}{d\sigma^{pA}
\over dx_F}(x_F/\alpha),
\ee
where $\alpha={x_F \over x_F+\Delta x_F}$, and $\Delta x_F$ is the shift of
\xf~due to the energy loss.
Since nuclear
shadowing effect can also contribute to the quarkonium production suppression
at large \xf,
in our calculation a value of $dE/dZ \simeq 0.5~ GeV/fm$ is found to
be able to fit the data quite well, which is consistent with the value obtained
from other considerations \cite{bro93}.

The comover contributions are included in the second exponential in
eq.(~\ref{crosseq}). We take the effective proper time $\tau_f=r_0/c_s$,
where $r_0$ is the projectile radius which we take to be $1.2~fm$,
and $c_s \sim 1/\sqrt{3}$ \cite{vog91}.

The results of our calculations are compared to E772/E789 data for
\jp~production at 800 $GeV/c$ in Figs.2 and 3, with different sets of
$\sigma_{co}$ and $\tau_0$. We find that
$\sigma_{co}=4.0~mb$ and $\tau_0=0.8~fm/c$ is the best fit, which corresponds
to a comover density with maximal value about $6~fm^{-3}$.
Fig.2 shows that a lower density (larger $\tau_0$)
 is not sufficient to explain the data. In Fig.3 we see that a
comover absorption cross section $\sigma_{co}$ smaller than $4.0~mb$ is
also insufficient to fit the data. Note that the dot-dashed curve in
Fig.3 corresponds to the case without comover contributions. The curve
illustrates a very small contribution of nulear absorption at low \xf,
as we expected. One can find in Fig.3 that
although the number of comovers is not very large in $p-A$ reactions,
comover interactions have a
much more important effect than nuclear absorption.

A comparison of our calculations for \upsln~production
with the E772 data is shown in Fig.4.
The dashed curve with
$\sigma_{co}=4.0~mb$ and $\tau_0=0.8~fm/c$
is somehow lower than the experimental points at $x_F \simeq 0.3$.
This may imply our over-estimate of comover contributions when
we fit the data at low \xf~region.

In the literature studies of nucleus-nucleus reactions indicated that
the comover density at an early stage of the collision is very high,
possibly from 1 to 5 $fm^{-3}$. In our work,
a comover density with a maximum$\sim 6~fm^{-3}$ is used when we fit the
E772/E789 data at low \xf.
With such a high density one would rather think of these
comovers as light quarks and antiquarks than as pions and low-mass
resonances \cite{gav90}. However, whether it is possible to obtain
such a high density in proton-nucleus collisions appears questionable.
Presumably some new mechanism might also contribute at
low \xf~region.

According to ref.\cite{kha94},
absorption in
confined hadronic matter is excluded as a possible cause of the \jp~suppression
observed in nucleus-nucleus collisions.
Since at low \xf~region neither energy loss nor
nuclear shadowing effect is important, it seems difficult to understand
the low \xf~suppression in proton-nucleus collisions, in which no
deconfined matter is expected.
Our work shows that
comover interactions play an important role at the low \xf~region,
and a comover absorption
cross section as large as $4~mb$ is needed
to confront the E772/E789 data.
In addition, a recent work observed that nuclear
antishadowing leads to enhanced \jp~production at low \xf, which is in
contradiction with the E772/E789 data \cite{liu94}.
These issues suggest
that the low \xf~suppression in $p-A$ reactions deserves careful studies,
and possibly requires some new mechanism.

To summarize, we have reproduced the overall \xf~dependence of quarkonium
production in proton-nucleus collisions, by combining comover contributions,
nuclear shadowing effect, nuclear absorption and energy loss mechanism
together. In particular, by including the \xf~dependence of comover
distribution
we show that comover interactions have a very important effect
when one attempts to explain the observed quarkonium suppression,
especially at low \xf~region, in proton-
nucleus collisions. However, our work indicates that a fairly high density
of comovers is needed to explain the E772/E789 low \xf~suppression, which
might imply some new mechanism occurs at this region as well.

\section*{Acknowledgments}
The authors would like to thank Q.H. Zhang and H.M. Hu for helpful discussions.

\newpage
\begin{center}
{\bf Figure Captions}
\end{center}

Fig.1. \xf~dependence of comover densities for different sets of $f$
and $\tau_0$.

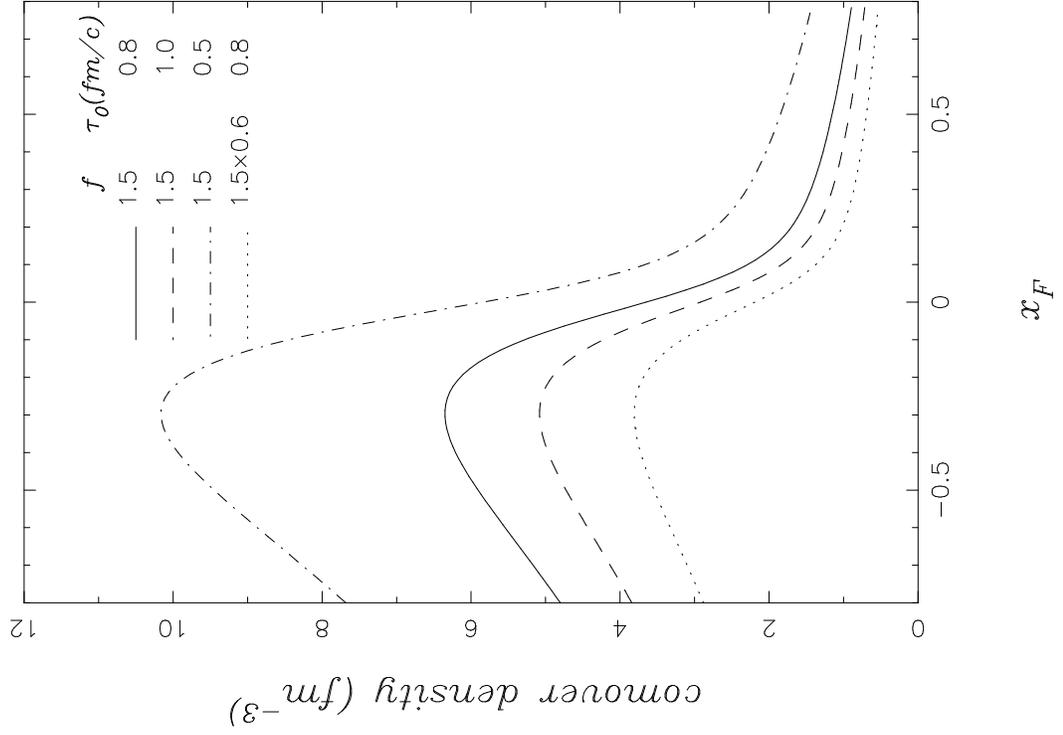
Fig.2. \xf~dependence for \jp~production from E772/E789 [3,4].
The curves compare our calculations to the data, with a fixed $\sigma_{co}=4.0
{}~mb$ and various values of $\tau_0$.

Fig.3. \xf~dependence for \jp~production from E772/E789 [3,4].
The curves compare our calculations to the data, with a fixed $\tau_0=0.8~fm/c$
and $\sigma_{co}=4.0
{}~mb,~2.0~mb$. The dot-dashed line corresponds to the case without comover
contributions.

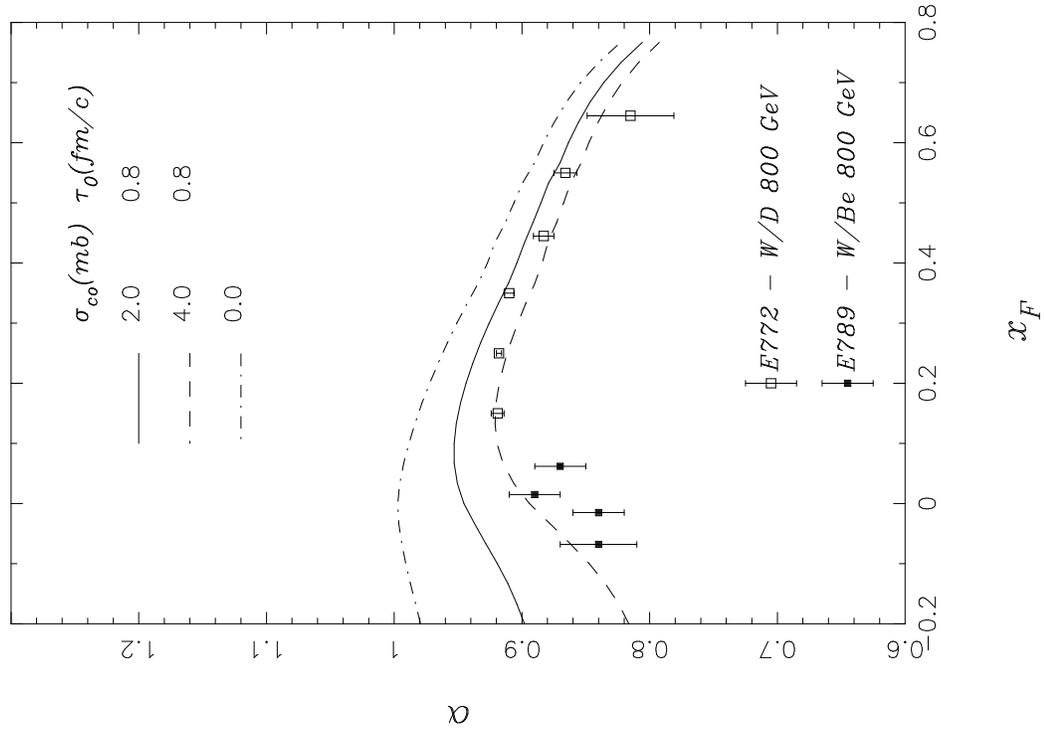
Fig.4. Our calculations of \xf~dependence for bottomonium production
with different sets of $\sigma_{co}$
and $\tau_0$
are compared to the E772 data [3].

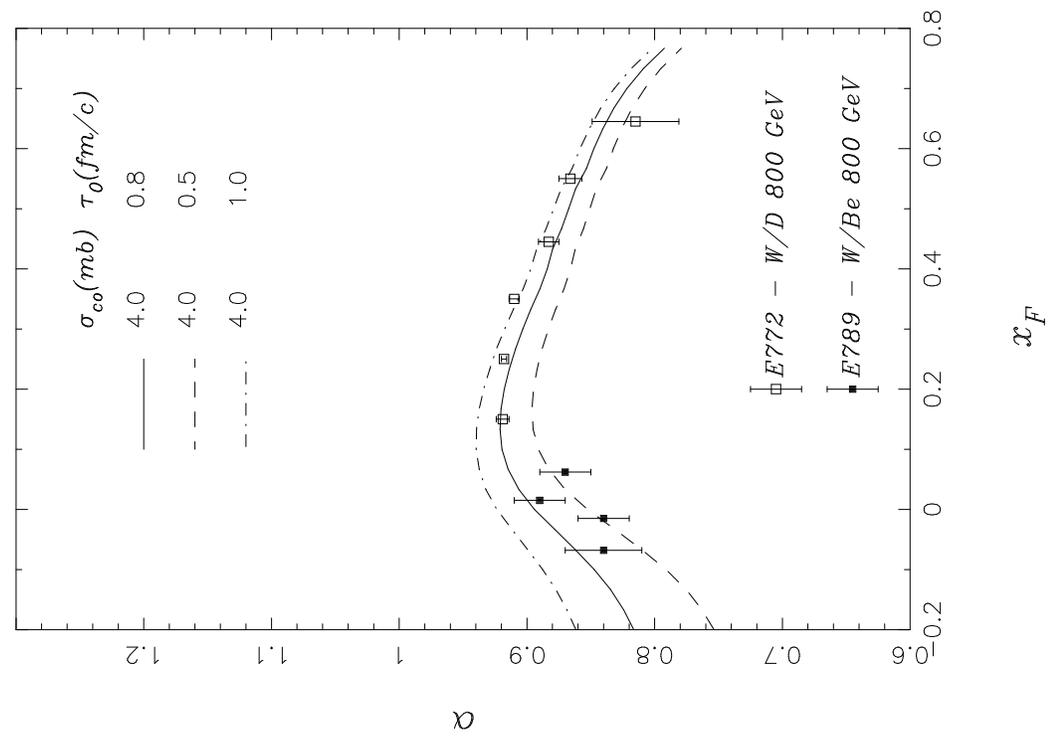
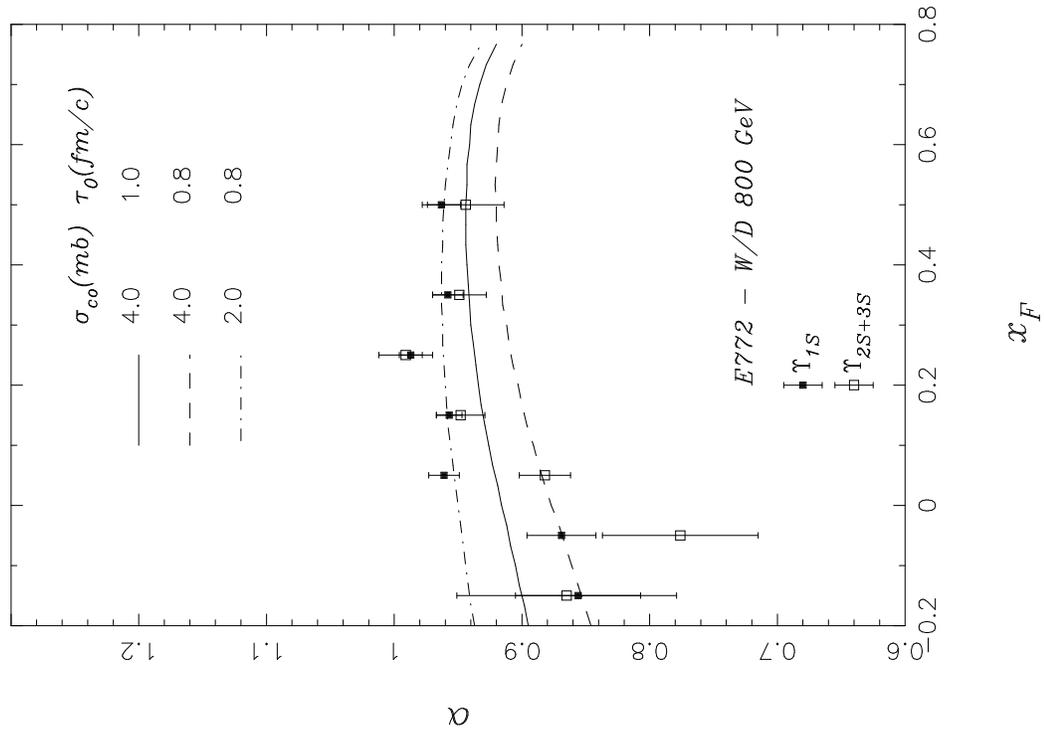
\end{document}